# Enhanced pyroelectric and piezoelectric properties of PZT with aligned porosity for energy harvesting applications

Yan Zhang [a], Mengying Xie [a], James Roscow [a], Yinxiang Bao [b], Kechao Zhou [b], Dou Zhang [b, *], Chris R. Bowen [a, *]

This paper demonstrates the significant benefits of exploiting highly aligned porosity in piezoelectric and pyroelectric materials for improved energy harvesting performance. Porous lead zirconate (PZT) ceramics with aligned pore channels and varying fractions of porosity were manufactured in a water-based suspension using freeze casting. The aligned porous PZT ceramics were characterized in detail for both piezoelectric and pyroelectric properties and their energy harvesting performance figures of merit were assessed parallel and perpendicular to the freezing direction. As a result of the introduction of porosity into the ceramic microstrucutre, high piezoelectric and pyroelectric harvesting figures of merits were achieved for porous freeze-cast PZT compared to dense PZT due to the reduced permittivity and volume specific heat capacity. Experimental results were compared to parallel and series analytical models with good agreement and the PZT with porosity aligned parallel to the freezing direction exhibited the highest piezoelectric and pyroelectric harvesting response; this was a result of the enhanced interconnectivity of the ferroelectric material along the poling direction and reduced fraction of unpoled material that leads to a higher polarization. A complete thermal energy harvesting system, composed of an parallel-aligned PZT harvester element and an AC/DC converter successfully demonstrated by charging a storage capacitor. The maximum energy density generated by the 60 vol.% porous parallel-connected PZT when subjected to thermal oscillations was 1653 $\mu J/cm^3$ respectively, which was 374% higher than that of the dense PZT with an energy density of 446 $\mu J/cm^3$. The results are of benefit for the design and manufacture of high performance porous pyroelectric and piezoelectric materials in devices for energy harvesting and sensor applications.

# 1. Introduction

Energy scavenging, or energy harvesting, which involves the conversion of ambient energy sources of energy such as mechanical vibrations, heat and light in useful electrical energy continues to receive significant industrial and academic interest. This stems from the ability of harvesting systems to provide a route for the realization of autonomous and self-powered low-power electronic devices[1-3]. Among the wide variety of potential sources of ambient energy, the harvesting of energy from vibrations, movement, sound and heat are considered to be highly promising energy harvesting technologies [4-6].

Mechanical waste energy that is generated by a vibrating structure or a moving object, is a ubiquitous form of energy that can be harvested from our surroundings[7]. The three common mechanisms for the conversion of vibrations into electricity include electromagnetic, electrostatic and piezoelectric techniques[8, 9]. Piezoelectric materials have significant potential due to their high power density, and adaptability in terms of the variety of macro- and micro-scale fabrication methods[10, 11]. In addition to mechanical energy, waste heat is a necessary by-product of all thermodynamic cycles implemented in power, refrigeration, and heat pump processes [12] and recovering even a fraction of this energy has the potential for an economic and environmental impact. The conversion of heat directly into electricity can be achieved by thermoelectricity[13] or pyroelectricity[14]. The advantages of pyroelectric materials, compared to thermoelectrics, is that they do not require bulky heat sinks to maintain a temperature gradient, and have the ability to operate with a high thermodynamic efficiency for converting temperature fluctuations into useable electrical power[15]. Energy conversion by heating and cooling a pyroelectric material therefore offers a novel way to convert heat into electricity and the approach has attracted interest in application areas such as low-power electronics and battery-less wireless sensors[16]. Since all pyroelectric materials are piezoelectric it is of significant interest to utilize piezoelectric and pyroelectric materials for energy harvesting applications with electrical properties that are readily adjustable and tailorable to suit the harvestable energy source and are also sufficiently robust to survive any applied mechanical loads and thermal strains.

In order to assess the properties of materials for energy harvesting a variety of performance *figures of merit* containing combinations of physical properties have been developed to describe the ability of materials to generate energy for practical applications. For piezoelectric and pyroelectric energy harvesting applications, the following figures of merit have been widely used for materials selection and materials design[17-20]:

$$FoM_{ij} = \frac{d_{ij}^2}{\varepsilon_0 \varepsilon_{33}^T} \quad (1)$$

$$F'_E = \frac{p^2}{\varepsilon_0 \varepsilon_{33}^\sigma \times (C_E)^2} \quad (2)$$

where $FoM_{ij}$ is a piezoelectric harvesting based figure of merit to assess the electrical energy density under a mechanical stress and $F'_E$ is a pyroelectric harvesting figure of merit indicating the energy density due to a temperature change. In Eqns. 1 and 2, $p$ is the pyroelectric coefficient, $\varepsilon$ is the relative permittivity (or dielectric constant), $\varepsilon_0$ is the permittivity of free space, $c_E$ is the volume specific heat capacity and $d_{ij}$ is the piezoelectric charge coefficient, where $d_{33}$ and $d_{31}$ are the longitudinal and transverse piezoelectric charge coefficients respectively. These figures of merit take into account the piezoelectric, pyroelectric and thermal properties and therefore indicate which materials are likely to perform well in specific energy harvesting applications. Equations 1 and 2 indicate that the important requirements for improved performance are low relative permittivity, low specific heat capacity and high piezoelectric and pyroelectric coefficients.

The two figures of merit above indicate that a porous structure whose permittivity is reduced due to the introduction of low permittivity pores would have beneficial consequences for piezo- and pyro-electric energy harvesting applications; however, it should be noted that the benefit is *only* achieved if the piezoelectric or pyroelectric coefficients are not reduced significantly by the presence of the pores. In general, it has been considered that porous piezoelectric and pyroelectric materials containing pores can be considered as a porosity (air)-ceramic matrix composite[21, 22]. It is well known that the properties of a ferroelectric composite depend on the depolarization factor, the connectivity of the phases and the permittivity ratio the two phases[23].

1.1 Porous piezoelectric composites with 2-2 connectivity

Laminated 2-2 connectivity based piezoelectric and pyroelectric composites, are of interest due to their simple architecture and their superior actuation and sensing abilities [24-26]. In the '2-2' system the first number describes the connectivity of the active PZT ceramic phase and the second refers to the passive matrix phase (in this case the pore volume). It has been demonstrated in our previous research [27] that PZT formed via ice-templating with highly aligned porosity exhibits a characteristic 2-2 connectivity along the freezing and poling directions. These materials exhibited a mechanical strength ~300% higher than conventional porous structures with randomly distributed porosity, and the pyroelectric figure of merits are 140%-360% times higher than the dense counterpart. Due to the unidirectional freezing nature of the ice-templating method, the resulting microstructure is highly oriented and strongly influences the corresponding physical, chemical and mechanical properties of the composite; these properties will be different in the directions that are parallel and perpendicular to the freezing direction, but are yet to be reported in detail for ferroelectric materials.

In this work freeze casting, also termed ice-templating, was utilized to fabricate porous PZT ceramics with tailored and aligned porosity to produce high performance materials for piezoelectric and pyroelectric energy harvesting applications.

The piezoelectric and pyroelectric parameters and figures of merit for the energy harvesting applications of the porous PZT ceramics both parallel and perpendicular to the freezing direction have been investigated in detail. Based on the properties of these highly aligned materials, the porous composites are then utilized in an energy harvester and are shown to generate more energy compared to the dense material. The work demonstrates that tailored porosity can generate improved materials for piezoelectric and pyroelectric energy harvesting devices.

## 2. Experimental

Commercial lead zirconate titanate powder (PZT-51, Zibo Yuhai Electronic Ceramic Co., Ltd, Zibo, China) with a mean particle size of ~3 μm was ball-milled for 48 h using a planetary ball mill to reduce the particle size $d_{50}$ to ~0.5 μm. Deionized water, polyvinyl alcohol (PVA, 420, Kuraray Co. Ltd, Japan) and ammonium polyacrylate (HydroDisper A160, Shenzhen Highrun Chemical Industry Co. Ltd, P. R. China) were used as the freezing vehicle, the binder and the dispersant, respectively. PZT suspensions with different solid loadings consisting of PZT powders, 1 wt.% dispersant and 1 wt.% PVA binder were ball-milled for 24 h in zirconia media and de-aired by stirring in a vacuum desiccator until complete removal of air bubbles. In the freeze casting process, the freezing operation is undertaken by a customized double-sided setup as described previously [28, 29]. The PZT suspensions were poured into a cylindrical polydimethylsiloxane (PDMS) mold (φ: 10 mm, H: 15 mm), which was then transported to a copper cold finger placed in a liquid nitrogen container. After solidification of the mixture was complete, the samples were freeze-dried for 24 h via ice sublimation in a vacuum chamber (~10 Pa) of a freeze-drier (FD-1A-50, Beijing Boyikang Medical Equipment Co., Beijing). The porous green bodies were placed in a furnace and heated in air with a heating rate of 1°C/min at 500 °C for 2 h to burn out the organic additives and sintered at 1200 °C for 2 h under a PbO-rich atmosphere, then finally allowed to cool down naturally in a closed furnace.

During the freeze casting process, dense and cellular zones are formed in the samples at the beginning of the freezing cycle, as shown in Fig. 1 (A). This region has a total thickness of less than 250 μm [30] and is removed by cutting the lower 0.5 mm of material via a low-speed precision diamond sectioning saw machine (Buehler, IsoMet LS). The freeze-cast porous samples of the same porosity (Fig. 1 (B)) were further cut both parallel (Fig. 1 (C)) and perpendicular (Fig. 1 (D)) to the freezing direction to assess their properties in different orientations with almost the same volume. The thickness, c, of both types of porous samples were ~ 1.0 mm. Samples cut parallel to the freezing direction had a length a= 14.63±0.43 mm and width b=5 mm. Samples cut perpendicular to the freezing direction has a diameter, φ= 9.65±0.14 mm.

The microstructures of the sintered samples, such as their pore structures and the structure of the PZT walls, were characterized using scanning electron microscopy (SEM, JSM-6480LV, JEOL Techniques, Tokyo, Japan). The apparent porosity of the composites was derived from density data obtained by the Archimedes's method[31]. The compressive strength of the sintered sample (diameter ~9.5 mm and height ~13 mm) was measured with a crosshead speed of 0.2 mm/min using an Electronic Universal Testing Machine (KD11−2, Shenzhen KEJALI Technology Co. Ltd., China). An average of five measurements was taken for each sample. The porosity of the porous ceramic was calculated using equation 3:

$$porosity = 1 - \frac{\rho_{porous\ PZT}}{\rho_{bulk}} \times 100\% \qquad (3)$$

where $\rho_{porous\ PZT}$ is the density of the porous sample, $\rho_{bulk}$ is the density of the bulk PZT.

Before electrical properties were measured, corona poling was conducted on the sintered dense and porous samples at 120 °C by applying a potential difference of 14 kV to a point source above the sample for 15 min, and samples were then aged for 24 h before testing. The longitudinal piezoelectric strain coefficient ($d_{33}$) and the transverse piezoelectric strain coefficient ($d_{31}$) were measured using a Berlincourt Piezometer (PM25, Take Control, UK). Measurements of the relative permittivity ($\varepsilon$) of the sintered materials were carried out from 1Hz to 1MHz at room temperature (~30ºC)

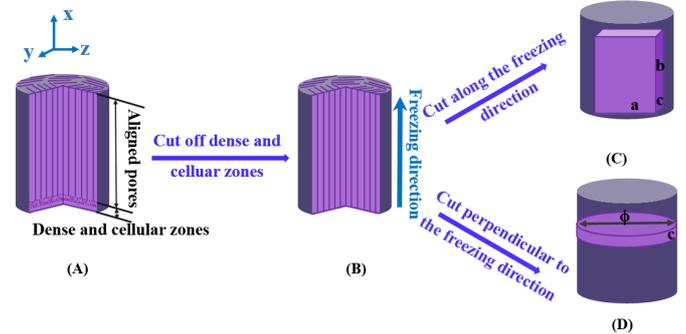

**Fig. 1** Schematic of the preparation of the freeze-cast samples. (A) sintered PZT ceramic, (B) aligned porous PZT sample after removing the dense and cellular zones, (C) cutting along the freezing direction and (D) perpendicular to the freezing direction.

using an impedance analyzer (Solartron 1260, Hampshire, UK).

Ferroelectric properties, such as the maximum polarization, remnant polarization and coercive field, were measured using a Radiant RT66B-HVi Ferroelectric Test system on unpoled materials. The specific heat capacity ($C_p$) of PZT composites with different porosities was measured from 20 to 100 °C by a MicroSC multicell calorimeter from Setaram, with the Calisto program to collect and process the data. The density of the solid PZT, $\rho$, is 7.6 g/cm$^3$ (according to the data sheet), therefore the volume specific heat, $C_V$, can be defined as:

$$C_V = \rho\ C_p \qquad (4)$$

The pyroelectric short circuit current ($I_P$) and pyroelectric open circuit voltage (V), were measured by an electrometer (Model 6517B, Keithley Instruments, Cleveland, OH), and the pyroelectric coefficient was determined by the Byer-Roundy method[32] and derived from:

$$I_P = p\, A\, \frac{dT}{dt} \qquad (5)$$

To demonstrate the effectiveness of the porous materials a thermal energy harvesting demonstrator was constructed, with a 1 µF charging capacitor used as an electrical storage element when the energy harvester was subjected to cyclic temperature fluctuations at 1.6 °C/sec. The temperature of the sample was continuously monitored using a K-type thermocouple with a response time of 0.5 s. A charging curve of the storage capacitor was measured for the range of materials, resulting in a maximum voltage on the storage capacitor without any electrical load.

## 3. Results and discussion

### 3.1 Microstructure and mechanical properties

As an example of a typical microstructure, Fig. 2 (A)-(D) shows SEM images of the microstructures of sintered freeze cast PZT ceramics in directions perpendicular (Fig. 2 (A)) and parallel (Fig. 2 (B)) to the unidirectional freezing direction for a composite with a porosity level of 40 vol.%. The presence of highly aligned and homogeneous lamellar pore channels with a pore width of 10-13 µm and a dense ceramic wall thickness of 5-8 µm can be observed in both directions, similar to our previous reports [28, 29]. The porous PZT ceramic parallel to the freezing direction exhibits a higher degree of parallelism compared to the perpendicular direction; compare Figs. 2 (A) and (B). In contrast to conventional isotropic porous structures with uniformly distributed porosity[33], a high fraction of ceramic links in the pore volume of the freeze-cast porous PZT is observed in both directions. These ceramic links are located in the adjacent lamellar walls of the long-range ordered pore channels, as shown by the red arrows of Fig. 2 (C) and (D) which shows higher magnification images of the microstructure. These ceramic bridges have been observed in a wide variety of freeze-cast porous ceramics which employ water-based suspensions [30], and for a variety of porosities ranging from 25 to 45 vol.% in the porous PZT ceramics in our previous work [27]. Under unidirectional freezing conditions, the growth velocity of the ice front along x- or y- axis in Fig. 1 was $10^2$-$10^3$ times faster than perpendicular to the freezing direction (the z-axis in Fig. 1)[30]. This results in aligned lamellar pore channels and dendritic ceramic bridges that are a replica of the orientated growth of the ice crystals.

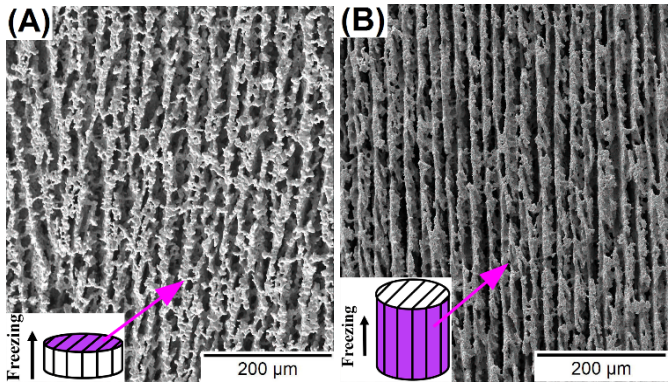

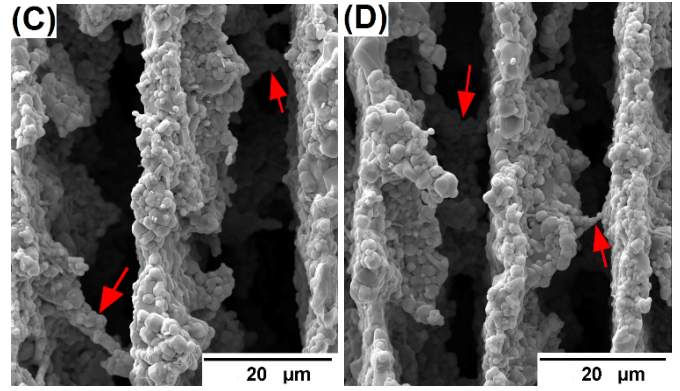

**Fig. 2** SEM images of the (A) parallel-connected (perpendicular to freezing direction) and (B) series-connected (along the freezing direction) porous PZT at a porosity of 40 vol.% with high magnification images in (C) and (D), respectively.

The compressive strengths of PZT ceramics with the porosities ranging from 20 to 60 vol.% in both the parallel and perpendicular directions to the unidirectional freezing direction has been shown in Fig. S1. Compared with the conventional porous PZT ceramics with uniformly distributed porosity[27], the PZT with pores perpendicular to the freezing direction exhibited a compressive strength of 320%-580% higher than that of the conventional porous PZT for porosities ranging from 20 to 60 vol.%, see Fig. S1. The strength along the freezing direction was ~200% higher than that of the uniformly distributed porous PZT. This high strength results from the unidirectional lamellar support and the strong bonding originating from the ceramic bridges between the lamellar ceramic walls, as seen in Fig. 2.

Fig. 4 shows a schematic diagram of the poling direction applied to the freeze-cast porous PZT that were cut both parallel (as in Fig. 1(C)) and perpendicular (as in Fig. 1(D)) to the freezing direction. Both types of PZT were explored to examine the influence of the pore orientation on poling behaviour and electrical properties, as shown in Fig. 4 (A) and (C). To model the influence of the composite microstructure, the highly aligned pores can be considered analogous to the classical parallel-connected (Fig. 4 (B)) and series-connected (Fig. 4 (D)) models used in estimating the piezoelectric and pyroelectric effects of diphasic composites, respectively [34].

### 3.2 Polarization-field (P-E) loops

Fig. 5 (A)(B) and (C)(D) show the P-E hysteresis loops of the parallel-connected and series-connected freeze cast PZT

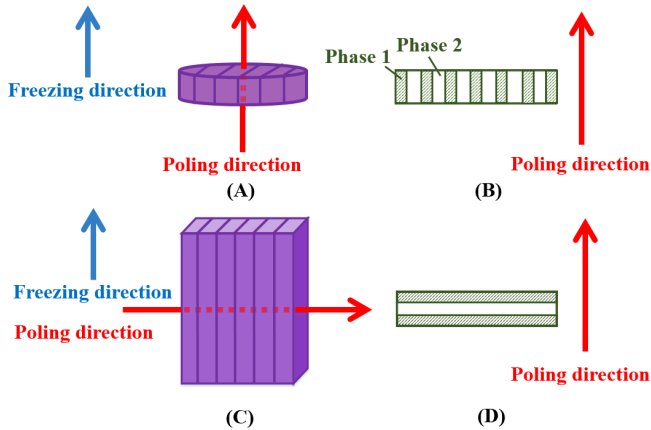

**Fig. 4** Schematic diagram of the poling processes conducted on (A) parallel-connected and (C) series-connected porous PZT, which can be considered analogous to the classical (B) parallel and (D) series models, respectively.

ceramics at different maximum electric fields and porosities, respectively. Fig. 5 (A) and (C) show, as an example, the development of the P-E loops with increasing maximum applied field for a 20 vol.% porous ceramic. The P-E loops become more square as the maximum applied field increases since the remnant polarization ($P_r$) and coercive field ($E_c$) increases. The $E_c$ increases rapidly at maximum fields between 9-12 kV/cm due to increased domain switching for the parallel-connected PZT and between 9-15 kV/cm for the series-connected PZT. The increase in $E_c$ with maximum field then slows down substantially at higher applied fields due to saturation (13-16 kV/cm for parallel-connected, 18-24 kV/cm for series-connected). Saturated and rectangular loops of the parallel-connected and series-connected freeze cast PZT ceramics were achieved at 16 kV/cm and 24 kV/cm respectively; higher electric fields than those presented here could not be applied due to dielectric breakdown.

Fig. 5 (B) and (D) show the parallel-connected and series-connected porous PZT ceramics with various porosities (20-60 vol.%), respectively. All materials exhibited a well-developed and almost symmetric hysteresis loops. The $P_r$ of the parallel-connected PZT decreased gradually from 16.5 to 6.4 µC/cm², and the $E_c$ monotonously increased from 7.7 to 9.1 kV/cm as the porosity fraction increased at a maximum electric field of 16 kV/cm. The $P_r$ of the series-connected PZT decreased from 11.3 to 3.5 µC/cm², and the $E_c$ increased from 8.9 to 10.3 kV/cm with increasing porosity at a maximum field of 24 kV/cm. Fig. S2 also shows for comparison the dense PZT with a $P_r$ of 35.0 µC/cm² and coercive field of 8.7 kV/cm.

For both types of porous PZT, the remnant polarization decreased gradually when the porosity increased from 20 to 30 vol.%, followed by a more rapid decrease when the porosity was higher than 40 vol.%. Increasing the porosity reduced the remnant polarisation since the porous PZT ceramics have proportionally reduced active material and ferroelectric domains compared with dense PZT, resulting in a lower polarisation. It should be noted that the parallel-connected PZT exhibited a more square P-E loop, higher polarisation and lower coercive field compared to the series-connected PZT for the same porosity, e.g. at 40 vol.% $P_r$ is 16.5 µC/cm² and $E_c$ is 8.25 kV/cm for the parallel-connected PZT and is 11.3 µC/cm² and 9.6 kV/cm for the series-connected PZT. This is due to the improved interconnectivity of the PZT material along the polarisation direction for the parallel-connected PZT, see Fig. 4.

While the decrease in remnant polarisation with increasing porosity is simply due to the reduction in polarisable material, the increase in coercive field with an increase in porosity is less obvious. The porous ceramics are effectively composites with a high permittivity ceramic phase and a low permittivity pore phase (air). Such a structure strongly influences the distribution of the electric field throughout the material microstructure during the poling process. It has been shown that for porous ferroelectric ceramics, the local electric field in the ceramic phase near the pore is much lower than the applied field. This inhomogeneous field distribution results in the existence of unpoled areas along the poling direction[35]. Consequently, the parallel-connected PZT exhibit less unpoled regions and easier domain switching and polarisation compared to the series-connected materials, leading to higher polarisation and lower coercive field.

A Finite Element Modelling approach has been used to investigate the effect of porosity in the freeze cast microstructure on the electric field distribution in the PZT during the poling process[36], and to account for the increase in coercive field with increasing porosity (see Fig 4(B) and 4(D)). Previous work[36] has demonstrated the ability of such models to predict the behaviour of porous PZT, although they have focused on uniformly distributed porosity within a ferroelectric material, i.e. systems with 3-0 and 3-3 connectivity. This approach has been adapted in this work to investigate the 2-2 connectivity structures achieved using the freeze casting process. A three dimensional cubic mesh of 29,791 (i.e. $31^3$) elements was split into an idealised 2-2 structure where alternate planes were assigned the properties of either unpoled PZT ($\varepsilon_r = 1585$[36]) or air ($\varepsilon_r = 1$). In reality, the freeze casting process does not produce ideal 2-2 structures due to ceramic links, as seen in Fig 2(C) and 2(D), and therefore a further step was incorporated in which randomly selected elements in the pore channels were reassigned the properties of PZT and randomly selected elements in the PZT channels reassigned the properties of air. To create geometries of varying porosity (20, 30, 40 and 50 vol.%) the ratio of PZT channel width to pore channel width was varied. To model the electric field distribution in the dense material (96% of theoretical density), 4% of elements were randomly assigned the properties of air with the remaining elements assigned the properties of unpoled PZT. An electric field was applied across the structure parallel to the lamellar channel direction and the local electric field in each element was recorded.

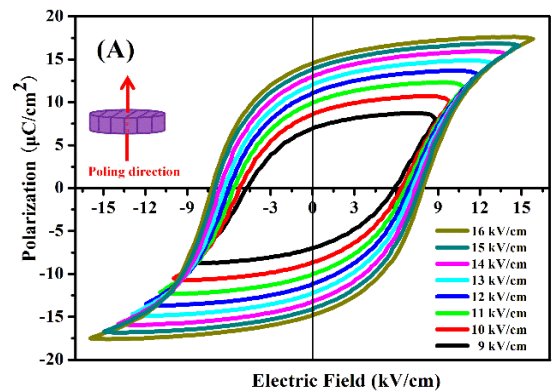

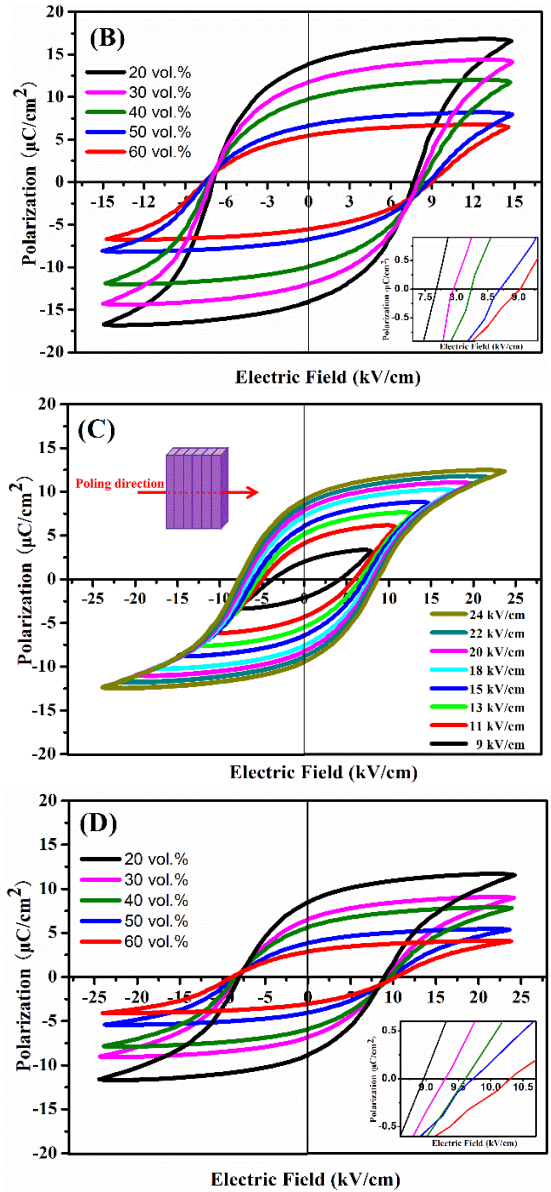

**Fig. 5** P-E hysteresis loops of the parallel-connected and series-connected freeze cast PZT. (A) parallel-connected and (C) series-connected PZT with the porosity of 20 vol.% and different electric fields, (B) parallel-connected and (D) series-connected PZT with the electric field of 16 kV/cm and 24 kV/cm respectively with different porosities. Insets in (B) and (D) are the coercive fields of the porous freeze-cast PZTs.

Fig. 6 shows the electric field distribution for dense PZT (4 vol.% porosity) and ceramics with increasing porosity levels. A high electric field concentration is observed within the low permittivity pore volume, with regions of low electric field observed in the immediate vicinity of the pore parallel to the direction of the applied field, which in a ferroelectric material may lead to incomplete poling in these regions, as shown in Fig. 6. Meanwhile, a lower field is observed in the ceramic regions, compared to the lower permittivity pore channels. A higher external electric field must therefore be applied to switch domains in the ceramic regions and hence the coercive field increases with increasing porosity, as in Fig 5(D). This phenomenon can be described by an adaptation of Gauss' law where the electric field, $E_f$, is related to the relative permittivity by:

$$E_f = \frac{q}{A \cdot \varepsilon_r \varepsilon_0} \qquad (6)$$

where $q$ is charge, $A$ is area and $\varepsilon_0$ is the permittivity of free space. In the dense ceramic, the electric field is relatively homogenous (see Fig. 6). However, in the porous structures the concentration of field in the low-permittivity pores result in the electric field no longer having a continuous and, linear path. As the porosity level increases it can be seen in Fig. 6 that the electric field in the ceramic moves to the lower magnitudes and is distributed over a wider distribution of electric fields. This results in an increase in coercive field with increasing porosity (as in Fig. 5(B) and 5(D)) since a larger external electric field must be applied to achieve domain switching.

### 3.3 Piezoelectric and dielectric properties

Fig. 7 (A) and (B) show the longitudinal ($d_{33}$) and transverse ($d_{31}$) piezoelectric charge coefficients of the series-connected and the parallel-connected freeze-cast PZT ceramics, along with the dense material. As shown in Fig. 7, the piezoelectric $d_{33}$ and $d_{31}$ charge coefficients decreased almost linearly with an increase of the porosity from 20 to 60 vol.%, similar to the results from Piazza et al.[37] who observed a decrease of $d_{33}$ in the porosity range from 4 vol.% to 45 vol.% for porous PZT using graphene as the pore-forming agent. This behaviour is due to the decrease of the polarisation of the material after the introduction of the porosity, as seen in the P-E loops in Fig. 5 (B) and (D). Moreover, the presence of concentrated stress can exist in the irregular dendritic ceramic area on the surface of the aligned pores which can be also responsible for the decrease of piezoelectric coefficients compared with the dense

PZT. The $d_{33}$ and $d_{31}$ of the parallel-connected porous PZT decreased by 16-35% and 15-58% respectively compared with the dense PZT, while there was a larger decrease in $d_{33}$ and $d_{31}$ of for the series-connected PZT (60-90% and 53-88%).

In addition, the parallel-connected porous PZT exhibited a longitudinal piezoelectric coefficient $d_{33}$ of 3.0 to 4.7 times higher than that of the series-connected samples, shown in Fig. 7(A), while the $d_{31}$ was ~2.5 times higher than that of the series-modelled PZT, shown in Fig. 7(B). These results are consistent with the P-E loops of Fig. 5 with a higher polarization in the parallel-connected PZT due the better interconnectivity of the PZT along the poling direction.

In the 2-2 connectivity composite, where the first phase is the active PZT piezoelectric ceramic, and the second phase is the passive pore channel (denoted as 'pc', henceforth), the piezoelectric coefficients for the series and parallel connection can be calculated by Equations S1-S4[34].

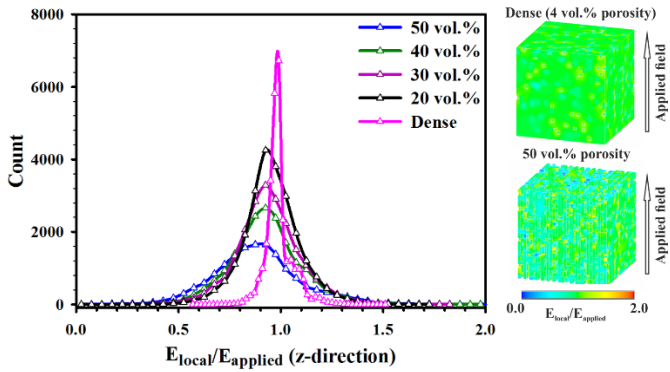

**Fig. 6** Model of electric field distribution in porous PZT and for dense (4 vol.% porosity) to 50 vol.% obtained by the Finite Element Modelling approach.

As can be seen from the SEM micrographs in Fig. 2, there are a number of ceramic bridges between the adjacent lamellar structures; therefore, in practice the aligned pore channel is a mixture of the PZT and air, rather than simply air. If we assume the piezoelectric coefficient and the elastic compliance of air are zero and infinite respectively, then the series and parallel equations can be simplified.

For series connection:

$$d_{33} = \frac{V^{PZT} d_{33}^{PZT}}{V^{PZT} + V^{pc} * m} \quad (7)$$

$$d_{31} = \frac{V^{PZT} d_{31}^{PZT}}{(V^{pc} + V^{pc}*m)*(V^{PZT} + V^{pc}*n)} \quad (8)$$

For parallel connection:

$$d_{33} = \frac{V^{PZT} d_{33}^{PZT}}{V^{PZT} + V^{pc}*n} \quad (9)$$

$$d_{31} = V^{PZT} d_{31}^{PZT} \quad (10)$$

where m is the ratio of the relative permittivity of the two phases ($\varepsilon_{33}^{PZT}/\varepsilon_{33}^{pc}$), while n is the ratio of the elastic compliances of the two phases ($s_{33}^{PZT}/s_{33}^{pc}$). The calculated results are the dotted lines and the experimental values are the data points in Fig. 7. After fitting with the simplified equations from (7) to (10), the fitting parameter, m, increased slightly from 6.012 to 6.089 in the series-connected freeze-cast samples with an increase of porosity, while the fitting parameter, n, decreased slowly from 0.781 to 0.755 in the parallel-connected porous PZT (see Tab. S1). Due to the small change of these two ratios, average vales of m=6.035 and n=0.763 were utilised in the fitting curve, shown in Fig. 7. From the SEM micrographs in our previous research, the amount of the ceramic links in the aligned pore channels reduced when the porosity increased, accounting for the reduction of the permittivity $\varepsilon_{33}^{pc}$ and the increase of the elastic compliance $s_{33}^{pc}$ of the aligned pore channels. Moreover, it was noticed that compared with a ratio of m~6.035 for the relative permittivity of the pore channel $\varepsilon_{33}^{pc}$ than that of the ceramic wall $\varepsilon_{33}^{PZT}$, the stiffness of the pore channel $1/s_{33}^{pc}$ was only 1.3 (=1/0.763) times lower that of the compliance of the lamellar ceramic wall $1/s_{33}^{PZT}$ in the $d_{33}$ of the parallel-connected and $d_{31}$ of the series-connected fitting values. As is known, nacre is a natural ceramic composite comprised of 95% calcium carbonate in the aragonite polymorph with ~5% organic macromolecules sandwiched in between. However, its work of fracture is ~3000 times greater than monolithic ceramics[38]. The aligned porous structures via freeze casting mimic the natural nacre structure due to the lamellar ceramic walls and the dendritic ceramic bridges were beneficial in stiffening the structure and limiting the torsion[39, 40], which may be responsible for the reduced stiffness ratio in both the parallel and series-connected porous PZT ceramics.

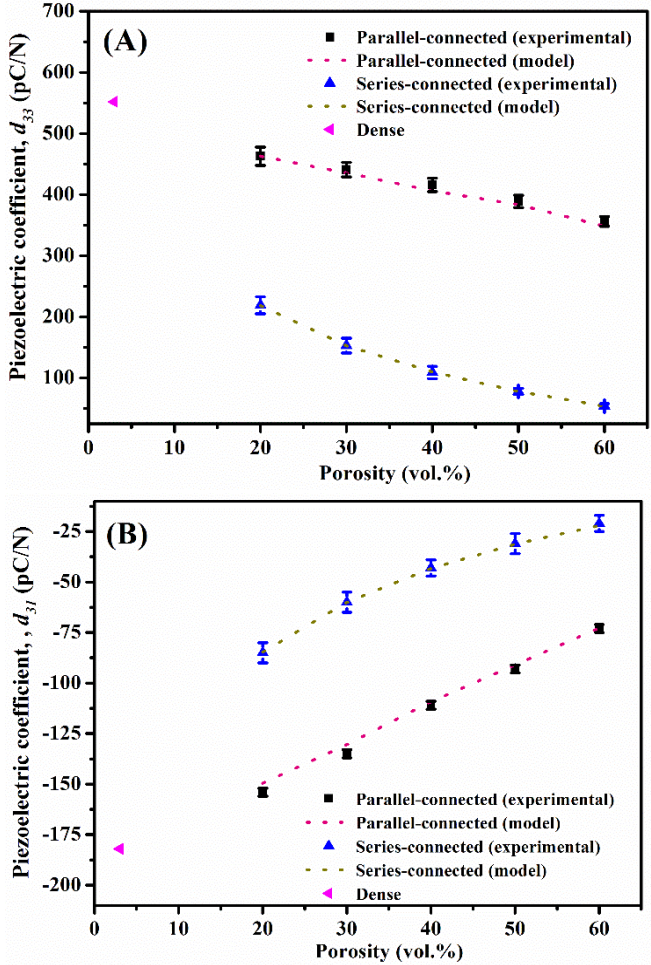

**Fig. 7** (A) Longitudinal ($d_{33}$) and (B) transverse ($d_{31}$) piezoelectric coefficients of porous freeze-cast PZT ceramics with both experimental and modelling results as a function of the porosity. Dense material also shown.

Fig. 8 shows the relative permittivity of the dense and freeze-cast porous PZT measured at 1 kHz and room temperature. The dielectric loss measured from 0.1-100 kHz are also shown in Fig. S3, with a similar loss for the ceramics across the porosity range. The relative permittivity decreased with increasing porosity in the range of 20-60 vol.% due to reduced volume fraction of high permittivity material, which is consistent with previously reported data [27] under different temperatures from 20 to 400 °C, showing that the dielectric loss of both the porous and dense PZT remained almost constant below the Curie temperature. Moreover, the parallel-connected freeze-cast PZT exhibited 3.7 - 41.3 times higher relative permittivity than the series-connected PZT on increasing the porosity from 20 to 60 vol.%, which is due to the improved interconnection in the parallel-connected freeze-cast PZT ceramic. In addition, it is noted that the relative permittivity of the parallel-connected freeze-cast PZT decreased almost linearly with porosity, while there was a non-linear relationship in series-connected PZT, in particular a gentle decrease can be observed with porosity >40 vol.%. Lichtenecker's equation can used to predict the dielectric function of a two-phase composite[41], given by Eqn. 11:

$$\varepsilon_L^k = v_1 \varepsilon_{PZT}^k + v_2 \varepsilon_{pc}^k \qquad (11)$$

where $\varepsilon_L^k$ is the composite relative permittivity, $v_1$ and $v_2$ are the volume fraction of PZT and pore channel in the porous PZT respectively, $\varepsilon_{PZT}^k$ and $\varepsilon_{pc}^k$ are the relative permittivity of PZT and pore channel, and k varies within the range of [-1,1], where k = 1 for a fully parallel connection and k = -1 for a fully series connection. Fig. 8 shows the fitting curves of the dotted lines and Tab. S1 presents the corresponding fitting results. As shown in Fig. 8, both the parallel-connected and series-connected freeze-cast PZT have a good fit with Lichtenecker's equation at all range of the porosities. Theoretically, the parameter k can be utilised to describe a transition from series anisotropy to at k = -1 to parallel anisotropy at k = 1. [42] As shown in Tab. 1, the k in series-connected PZT decreased from -0.016 to -0.043 while in parallel-connected PZT the k value increased from 0.813 to 0.905 with an increase of porosity, demonstrating that both types of freeze-cast ceramics possessed an increased degree of anisotropy with increasing porosity. Meanwhile, the reduced number of the ceramic links in the aligned pore channels with high porosity can also be responsible for the higher degree of anisotropy of the freeze-cast ceramic.

Fig. 9 shows the piezoelectric generator figure of merit, $FoM_{ij}$, calculated by Eqn. (1) at a constant stress as a function of porosity. For the longitudinal $d_{33}$ piezoelectric figure of merit, the parallel-connected PZT exhibited a higher value than the series-connected PZT. However, the parallel-connected PZT had a lower transverse $d_{31}$ piezoelectric figure of merit ($FoM_{31}$) in the whole porosity range. In the series-connected PZT, both longitudinal $FoM_{33}$ and transverse $FoM_{31}$ piezoelectric figures of merit increase with increasing porosity and were higher than the dense material. The increase in the figures of merit with porosity in Fig. 9 results from the large decrease in relative permittivity with increasing porosity in Fig. 8. However, in the parallel-connected PZT, while the transverse $FoM_{31}$ figure of merit increased with an increase of porosity, the transverse $FoM_{31}$ figure of merit decreased with increasing porosity and was lower than the dense material when the porosity > 40 vol.%. This is due to the more rapid decrease of the $d_{31}$ (Fig. 7(B)) compared to the permittivity (Fig. 8) with porosity for this orientation. A further increase in the porosity level beyond 60 vol.% can result in a decrease of the piezoelectric figure of merit since $d_{33}$ falls rapidly at higher porosity [43]; in addition it has no benefit to the mechanical strength, as shown in Fig. S1.

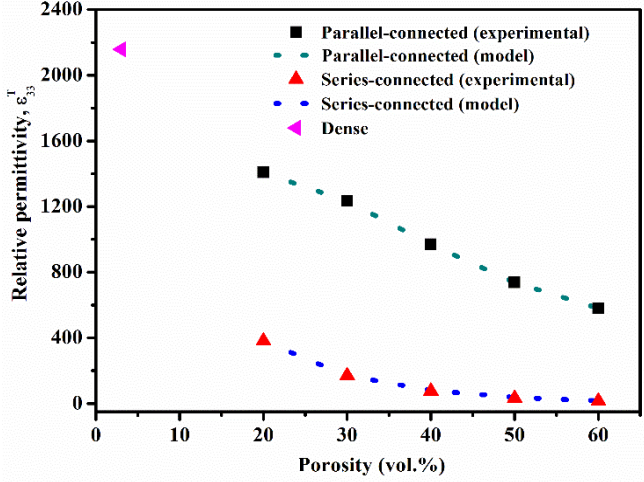

**Fig. 8** Relative permittivity of porous freeze-cast PZT ceramics with both experimental and modelling results as a function of the porosity. Dense material also shown.

**Tab. 2** Modelling parameters (k) of the relative permittivities of the freeze-cast porous PZT with both series and parallel connectivities.

| Porosity (%) | Series-connected | Parallel-connected |
|---|---|---|
| 20 | -0.016 | 0.511 |
| 30 | -0.037 | 0.611 |
| 40 | -0.038 | 0.632 |
| 50 | -0.039 | 0.639 |
| 60 | -0.043 | 0.692 |

### 3.4 Pyroelectric properties

Fig. 10 (A) and (B) show the pyroelectric coefficients and pyroelectric figure of merit ($F_E'$) of porous freeze-cast PZT ceramics. Theoretical pyroelectric formulations for the series and parallel connections are given by Equations (S5) and (S6) respectively. After simplifying these equations in a similar way to the above piezoelectric coefficients, equations (12) and (13) are obtained as below.

For series connection:

$$p = \frac{V^{PZT} p^{PZT}}{V^{PZT} + V^{pc} * m} \quad (12)$$

For parallel connection:

$$p = V^{PZT} p^{PZT} \quad (13)$$

Fig. 10(A) shows that the pyroelectric coefficient decreased in both types of the freeze-cast PZT ceramics with increasing porosity, and are lower than the dense PZT. The parallel-connected PZT possessed higher pyroelectric coefficient than that of the series-connected PZT at the same porosity, e.g. 3.0 µC/m²·K and 4.0 µC/m²·K for the series-connected PZT and parallel-connected PZT respectively at the 40 vol.% porosity. As seen in Fig. 5, the decreased polarization in the porous samples, results in the decreased pyroelectric coefficient (*p*) due to the relationship between pyroelectric coefficient and polarization where *p=dP/dT*, and *dP* and *dT* are the change in polarization and temperature respectively. The measured specific heat capacity of the porous PZT were 260, 227, 201, 163, 140 J/kg·K at room temperature when the porosity increased from 20 to 60 vol.%, shown in Fig. 10(B). The $F_E'$ figure of merit for thermal harvesting (Eqn. 2) are shown in Fig. 10(C). Although the pyroelectric coefficient of the parallel-connected PZT decreased by ~49%, see Fig. 10(A), the corresponding figure of merit $F_E' = p^2 / (\varepsilon_{33}^\sigma \times (C_E)^2)$ increased by ~405% when the porosity increased from 20 to 60 vol.% due to a combination of the reduced permittivity (Fig. 8) and specific heat capacity (Fig. 10(B)). The same trends were also found in the series-connected PZT with a larger decrease in the pyroelectric coefficient and lower increase in $F_E'$ which were ~76% and ~151%, respectively. The modelling results demonstrate a good fit with the experimental data.

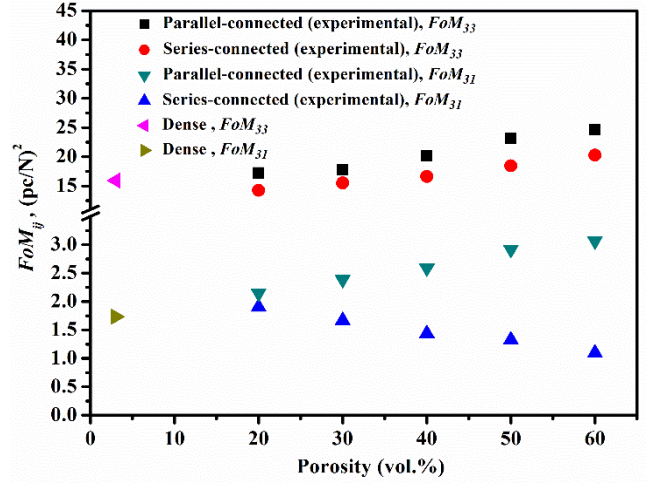

**Fig. 9** Piezoelectric $FoM_{33}$ and $FoM_{31}$ generator figure of merit at a constant stress of porous freeze-cast PZT ceramics as a function of the porosity. Dense material also shown.

In addition to potential energy harvesting applications, there are other important pyroelectric detection figures of merit to describe the performance of a material as a pyroelectric sensor [44]. These include the current responsivity, $F_I = p/C_E$, voltage responsivity, $F_V = p/(C_E \cdot \varepsilon_{33}^T \varepsilon_0)$ and signal to noise figure of merit, $F_D = p/(C_E \cdot (\varepsilon_{33}^T \varepsilon_0 \cdot tan\delta)^{0.5})$. The current responsivity, $F_I$, remains constant with increasing porosity for the parallel connected material since the decrease is pyroelectric coefficient is similar to the decrease in heat capacity, see Fig 11(A). For the series connected material there is a decrease in $F_I$, due to the low pyroelectric coefficient in this direction (Fig 10(A)). The voltage responsivity, $F_v$ in Fig 11(B), increases with increasing porosity in both directions due to the decrease in heat capacity and permittivity. The $F_V$ is larger for the series connected materials due to the lower permittivity, see Fig 8. Finally, $F_D$ increases with increasing porosity, Fig 11(C), in both directions due to reduced specific heat and permittivity; changes in loss are relatively small. Interestingly, while the series connection is advantageous for the $F_V$ and $F_D$ detection figures of merit, the parallel connection is highest for the $F_E'$ harvesting figure of merit, see Fig 10(C); this is due to its $p^2$ dependency (see Eqn. 2) and higher pyroelectric coefficient in this direction (Fig 10(A)).

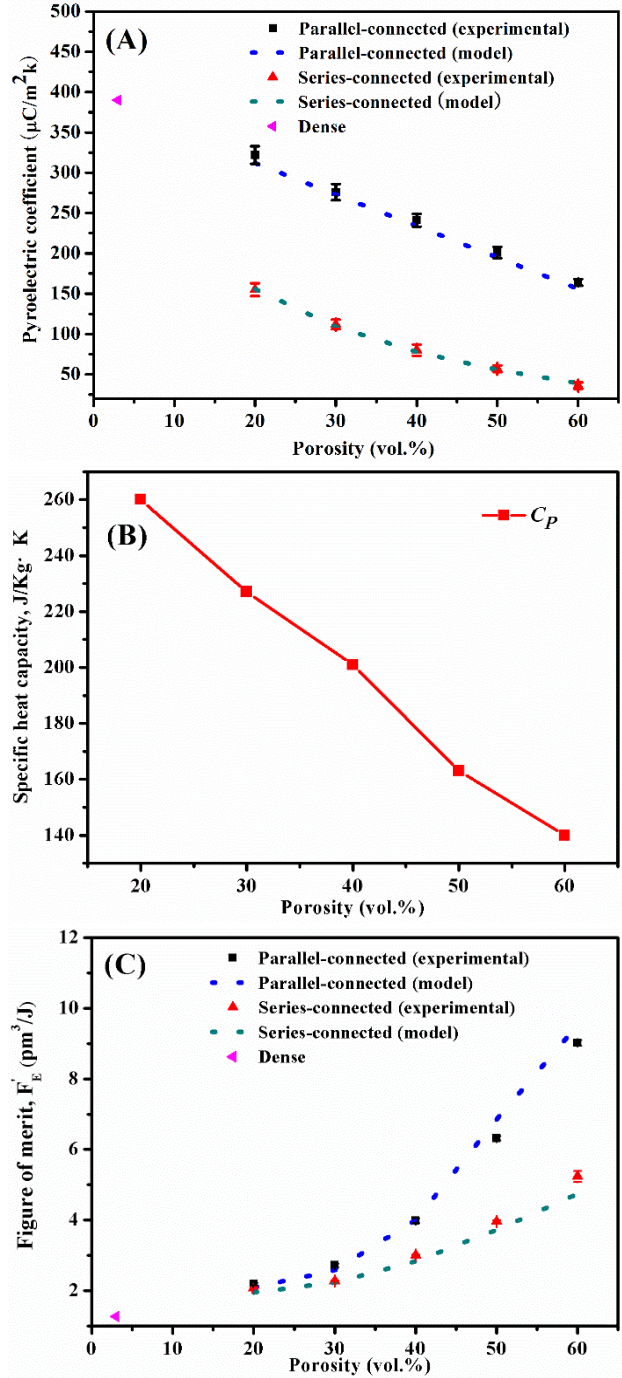

**Fig. 10** (A) Pyroelectric coefficients, (B) specific heat capacity and (C) pyroelectric figure of merit of porous freeze-cast PZT ceramics with both experimental and modelling results as a function of the porosity. Dense material also shown.

### 3.5 Pyroelectric energy harvesting demonstrator

As a consequence of the reduced permittivity (Fig 8) and volume specific heat capacity (Fig 10(B)) it can be seen that the $F_E'$ pyroelectric harvesting figure of merit increased with an increase of the porosity (as in Fig 10(C)) despite the decreasing pyroelectric coefficients. Furthermore, in order to demonstrate the potential of porous materials with improved energy harvesting figures of merit a demonstrator system for harvesting thermal fluctuations was developed. Fig. 11(A)-(D) shows the pyroelectric energy harvesting system, a schematic of the full bridge rectifier circuit with the AC-DC rectifier and storage capacitor, and the voltage level of the capacitor along with the generated electric energy/energy density with time during charging the capacitor as the pyroelectric element is subjected to temperature oscillations. In this work, an electrometer was used to measure the voltage of the charging capacitor generated by the porous PZT with parallel-connected porosity due to the improved figures of merit with increasing porosity, as seen Fig. 10(C). A Philips HP3631/01 IR lamp, with a power of 175 W, was employed as heat source to heating and cooling the PZT elements. Specifically, in a thermal energy harvesting application, the AC signals generated by the energy harvesting materials (both dense and the aligned porous PZT ceramics) must be rectified and stored in a capacitor. A full wave rectifier circuit similar to our previous report [45] is shown in Fig. 11(B), where the pyroelectric element is considered as a charge source with a shunt capacitor and resistor. The magnitude of the current / voltage varies with the porosity level of the piezoelectric element at the same temperature fluctuation.

The measured forward voltage drop of the rectifier circuit was 2V and during the charging process, the energy is $CV^2/2$, where $C$ and $V$ were the capacitance and the voltage of the capacitor, respectively. It can be clearly seen that on increasing the level of aligned porosity from 20 vol.% to 60 vol.%, the parallel-connected porous PZT produced an increasing peak voltage from 8.3 to 15.2 V after 3600s (Fig. 11(C)) with a corresponding increase of energy from 34 μJ to 116 μJ (Fig. 11 (D)) and energy density and from 490 to 1653 μJ/cm$^3$ (inset in Fig. 11 (D)). The dense PZT had the lowest voltage of 7.8 V, energy (31 μJ) and energy density (446 μJ/cm$^3$). Both the voltage and the energy density increased with an increase of the porosity level, which is entirely consistent with the improvement pyroelectric figure of merit shown in Fig. 10(B). A maximum voltage of 15.2 V was obtained in the storage capacitor utilising parallel-connected PZT with the highest porosity of 60 vol.% which demonstrated the fastest charging speed to obtain a stable voltage level of the capacitor, and the corresponding maximum energy and energy density (1653 µJ/cm³) was 374% higher than that of the dense PZT (value).

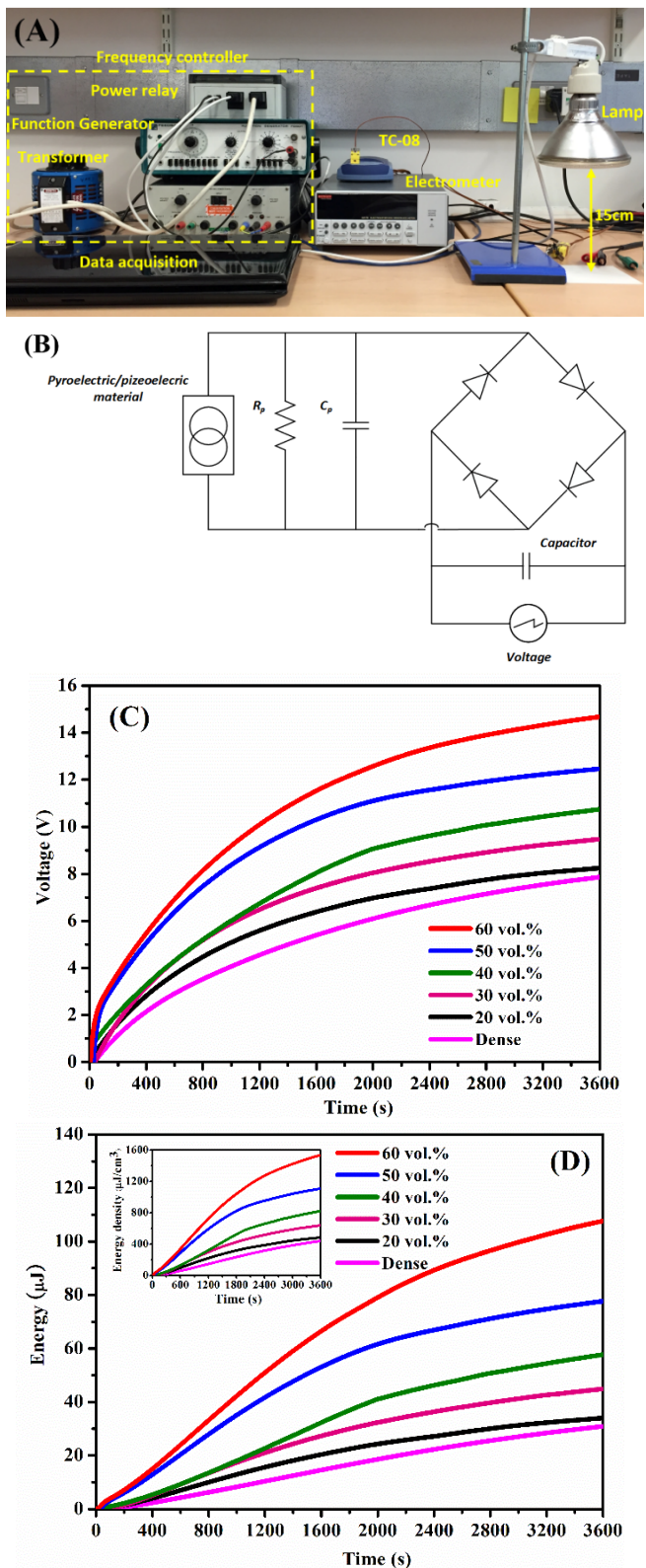

**Fig. 11** (A) The test setup platform of the pyroelectric energy harvester. (B) A schematic diagram of a pyroelectric power generator with circuitry. (C) Charging voltage vs time plots of the dense and parallel-connected porous PZT with various porosities ranging from 20 to 60 vol.% and the dense material. (D) Charging energy of the capacitor, the inset shows the charging energy density.

## Conclusions

This paper has demonstrated the significant benefits of using porosity to enhance the piezoelectric and pyroelectric properties of materials for energy harvesting. Novel 2-2 connectivity PZT ceramics with aligned pore channels have been prepared by freeze casting water-based PZT suspensions, with porosity levels ranging from 20 to 60 vol.%. The unidirectional lamellar microstructure and the strong lamellar bonding originating from the ceramic bridges between the lamellar walls were shown to be beneficial to achieve a higher compressive strength compared with the conventional porous PZT with uniformly distributed porosity. The change in remnant polarization and coercive field with porosity level is described and explained with modelling of the electric field distribution within the two-phase microstructure. The parallel-connected porous PZT exhibited a 300% to 470% higher longitudinal piezoelectric coefficient $d_{33}$ and ~250% times higher transverse piezoelectric coefficient $d_{31}$ than the series-connected PZT in the complete porosities range of 20 to 60 vol.%. As a result of the reduced permittivity and reduced specific heat capacity, the aligned porous PZT demonstrated excellent piezoelectric and pyroelectric figures of merit, compared with the dense counterpart. The modelling results utilising the coupling of the ratio of the dielectric constant and elastic compliance between the ceramic wall and pore channel showed good agreement with the experiment results. To demonstrate the improved performance of the aligned porous materials a pyroelectric energy harvesting device was constructed and clearly demonstrated that the fastest charging speed, and maximum energy density was obtained when utilising the highest porosity of 60 vol.% for a parallel-connected PZT. The measured energy density was 374% higher than that of the dense PZT and demonstrate that porous materials are highly attractive materials for energy harvesting applications.

## Acknowledgements

Dr. Y. Zhang would like to acknowledge the European Commission's Marie Skłodowska-Curie Actions (MSCA), through the Marie Skłodowska-Curie Individual Fellowships (IF-EF) (H2020-MSCA-IF-2015-EF-703950-HEAPPs) under Horizon 2020. Prof. C. R. Bowen would like to acknowledge funding from the European Research Council under the European Union's Seventh Framework Programme (FP/2007–2013)/ERC Grant

Agreement no. 320963 on Novel Energy Materials, Engineering Science and Integrated Systems (NEMESIS). The authors are also thankful for the experimental supports from Ms Xi Yuan and Ms Fengdan Xue, and the helpful advices and inspiring discussions with Dr. Hamideh Khanbareh and Mr Marcin Krasny.

# Supplementary information

# Enhanced pyroelectric and piezoelectric properties of PZT with aligned porosity for energy harvesting applications

Yan Zhang [a], Mengying Xie [a], James Roscow [a], Kechao Zhou [b], Yinxiang Bao [b], Dou Zhang [b,*], Chris R. Bowen [a,*]

[a] Department of Mechanical Engineering, University of Bath, BA2 7AY, United Kingdom

[b] State Key Laboratory of Powder Metallurgy, Central South University, 410083, China

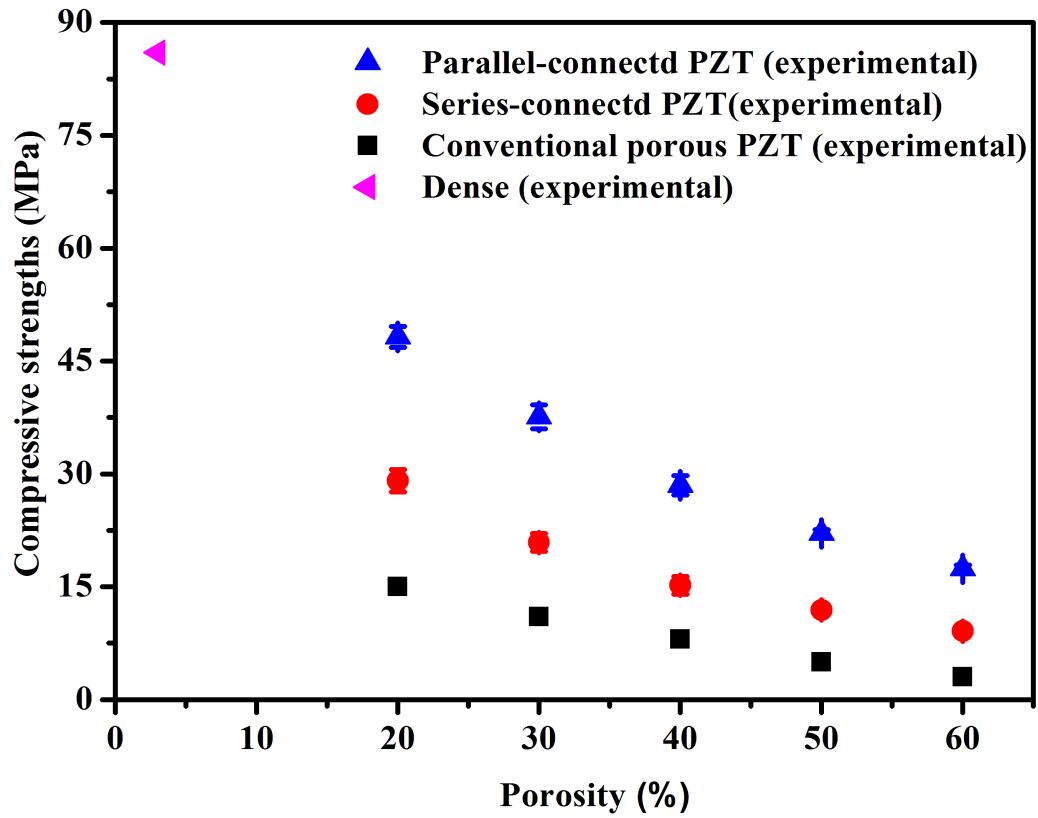

Figure S1 Compressive strengths of both parallel-connected and series-connected freeze-cast porous PZT. Data for conventional porous PZT with uniformly distributed porosity and dense PZT ceramics also shown.

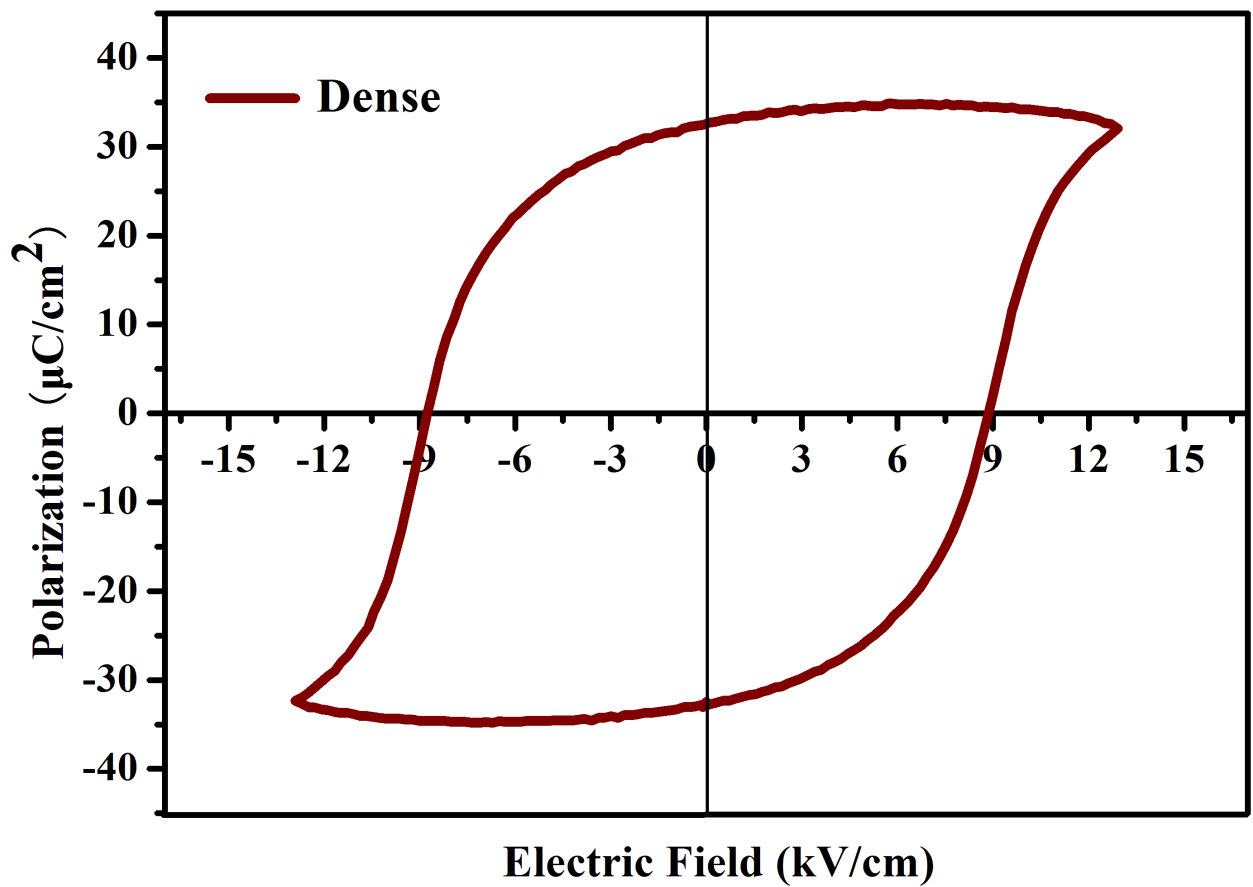

Figure S2 Polarisation (P) - electric field (E) hysteresis loops of the dense PZT.

## 3.3 Piezoelectric and dielectric properties

Piezoelectric coefficients for the series and parallel connection can be calculated by the following equations:

For series connection by Equations S1 and S2:

$$d_{33} = \frac{V^{PZT} d_{33}^{PZT} \varepsilon_{33}^{pc} + V^{pc} d_{33}^{pc} \varepsilon_{33}^{PZT}}{V^{PZT} \varepsilon_{33}^{pc} + V^{pc} \varepsilon_{33}^{PZT}} \quad (S1)$$

$$d_{31} = \frac{V^{PZT} d_{31}^{PZT} \varepsilon_{33}^{pc} s_{11}^{pc} + V^{pc} d_{31}^{pc} \varepsilon_{33}^{PZT} s_{11}^{PZT}}{(V^{PZT} \varepsilon_{33}^{pc} + V^{pc} \varepsilon_{33}^{PZT})(V^{PZT} s_{11}^{pc} + V^{pc} s_{11}^{PZT})} \quad (S2)$$

For parallel connection by Equations S3 and S4:

$$d_{33} = \frac{V^{PZT} d_{33}^{PZT} s_{33}^{pc} + V^{pc} d_{33}^{pc} s_{33}^{PZT}}{V^{PZT} s_{33}^{pc} + V^{pc} s_{33}^{PZT}} \quad (S3)$$

$$d_{31} = V^{PZT} d_{31}^{PZT} + V^{pc} d_{31}^{pc} \quad (S4)$$

## 3.4 Pyroelectric properties

Theoretical pyroelectric formulations for the series and parallel connections are given by Equations (S5) and (S6) respectively.

For series connection:

$$p = \frac{V^{PZT} p^{PZT} \varepsilon_{33}^{pc} + V^{pc} p^{pc} \varepsilon_{33}^{PZT}}{V^{PZT} \varepsilon_{33}^{pc} + V^{pc} \varepsilon_{33}^{PZT}} + \frac{2 V^{PZT} V^{pc} (\varepsilon_{33}^{pc} d_{33}^{PZT} - \varepsilon_{33}^{PZT} d_{33}^{pc})(\alpha^{pc} - \alpha^{PZT})}{(V^{PZT} \varepsilon_{33}^{pc} + V^{pc} \varepsilon_{33}^{PZT})[V^{PZT}(s_{11}^{pc} + s_{12}^{pc}) + V^{pc}(s_{11}^{PZT} + s_{12}^{PZT})]} \quad (S5)$$

For parallel connection:

$$p = V^{PZT} p^{PZT} + V^{pc} p^{pc} + \frac{V^{PZT} V^{pc} p^{PZT} (\alpha^{pc} - \alpha^{PZT})(d_{33}^{PZT} - d_{33}^{pc})}{V^{PZT} s_{33}^{pc} + V^{pc} s_{33}^{PZT}} \quad (S6)$$

where α is thermal expansion coefficient.

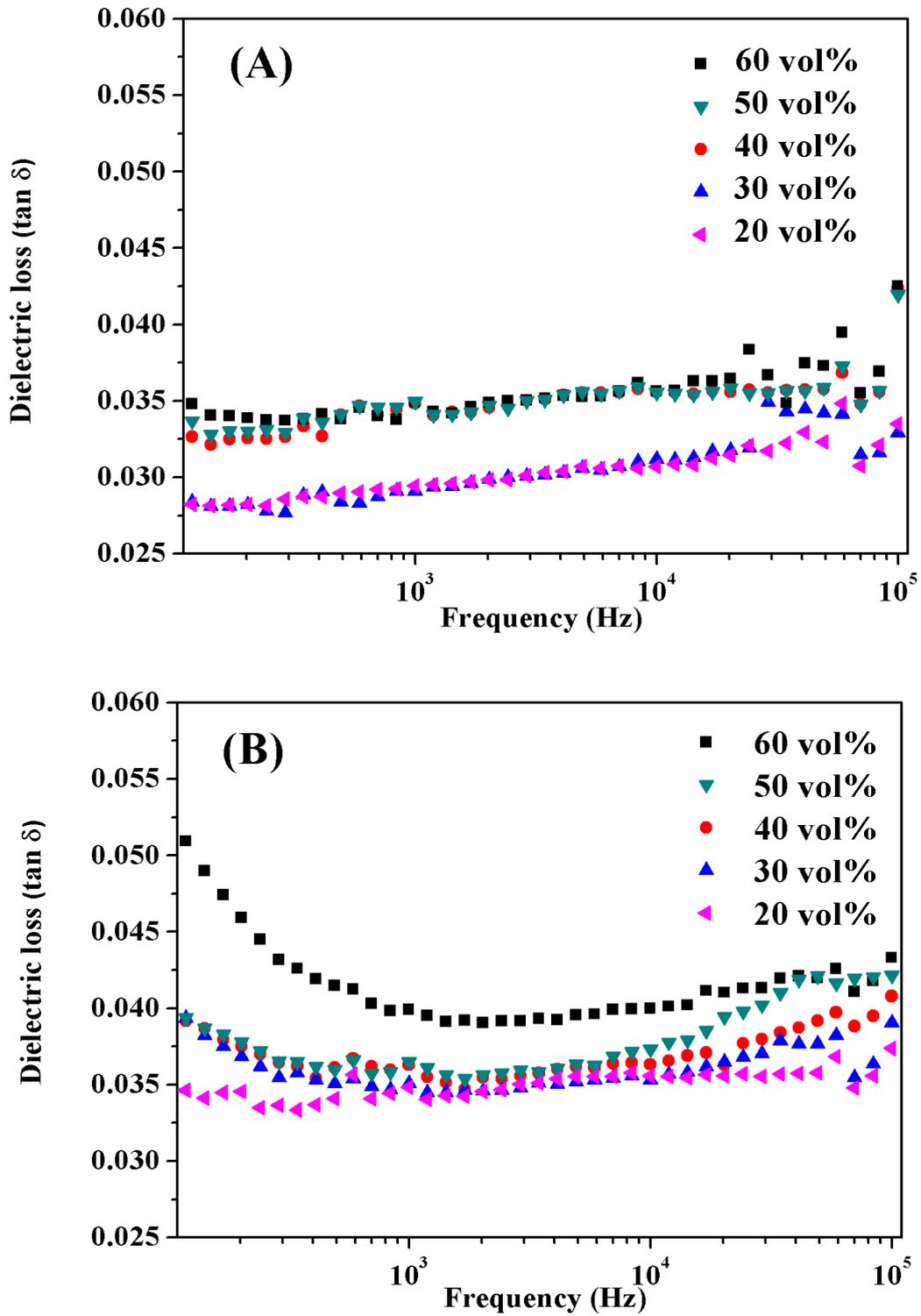

Figure S3 Dielectric loss of the aligned porous PZT with (A) Parallel-connected and (B) Series-connected modes.

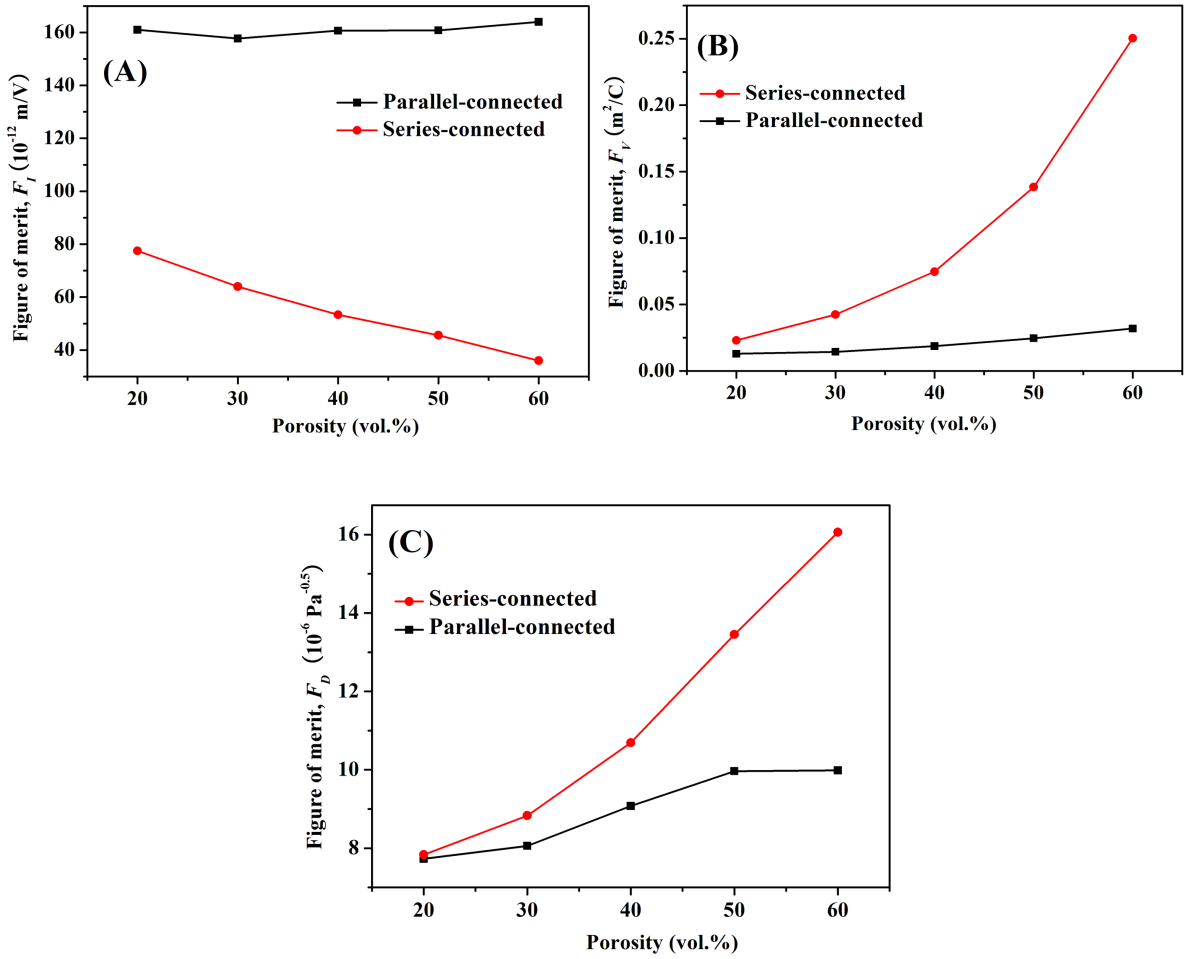

Figure S4 Figures of merit for the pyroelectric detectors. (A) Current responsivity $F_I$, (B) Voltage responsivity $F_V$ and (C) $F_D$.

Current responsivity is proportional to $F_I = \frac{P}{C_V}$, the voltage responsivity is proportional to $F_V = \frac{P}{C_V \varepsilon_r \varepsilon_0}$, in the case of a pyroelectric detector dominated by the AC Johnson noise, the detector figure of merit is proportional to $F_D = \frac{P}{C_V \sqrt{\varepsilon_r \varepsilon_0 \tan\delta}}$, where $p$ is the pyroelectric coefficient, $C_V$ is the volume specific heat capacity, $\varepsilon_r$ is the relative permittivity, $\varepsilon_0$ is the permittivity of free space.

Table S1 Model fitting parameters of the piezoelectric coefficients of the freeze-cast porous PZT with both series and parallel connectivities.

| Porosity (%) | $d_{33}$ | | $d_{31}$ | |
|---|---|---|---|---|
| | Series-connected | Parallel-connected | Series-modelled | Parallel-connected |
| | m | n | m, n | N/A |
| 20 | 6.012 | 0.781 | m=6.012, n=0.781 | - |
| 30 | 6.019 | 0.762 | m=6.019, n=0.762 | - |
| 40 | 6.021 | 0.759 | m=6.021, n=0.759 | - |
| 50 | 6.032 | 0.757 | m=6.032, n= 0.757 | - |
| 60 | 6.089 | 0.755 | m=6.089, n= 0.755 | - |